\documentclass[useAMS,usenatbib]{mn2e}
\usepackage{graphicx}
\usepackage{times}
\newcommand{\hompc}{\,h\,{\rm Mpc}^{-1}}
\newcommand{\mpcoh}{\,h^{-1}\,{\rm Mpc}}

\begin{document}

\title[Redshift-space distortions and projected clustering] {The effect of redshift-space distortions on projected 2-pt clustering measurements}

\author[Kelly Nock, Will J. Percival, Ashley J. Ross]{
\parbox{\textwidth}{
Kelly Nock\thanks{e-mail: kelly.nock@port.ac.uk (KN)}, 
Will J. Percival, Ashley J. Ross}
\vspace*{4pt} \\
Institute of Cosmology and Gravitation, University of Portsmouth,
Portsmouth, PO1 3FX, UK}

\date{\today} 
\pagerange{\pageref{firstpage}--\pageref{lastpage}} \pubyear{2008}
\maketitle
\label{firstpage}

\begin{abstract}
  Although redshift-space distortions only affect inferred distances
  and not angles, they still distort the projected angular clustering
  of galaxy samples selected using redshift dependent quantities. From
  an Eulerian view-point, this effect is caused by the apparent
  movement of galaxies into or out of the sample. From a Lagrangian
  view-point, we find that projecting the redshift-space overdensity
  field over a finite radial distance does not remove all the
  anisotropic distortions. We investigate this effect, showing that it
  strongly boosts the amplitude of clustering for narrow samples and
  can also reduce the significance of baryonic features in the
  correlation function.  We argue that the effect can be mitigated by
  binning in apparent galaxy pair-centre rather than galaxy position,
  and applying an upper limit to the radial galaxy separation. We
  demonstrate this approach, contrasting against standard top-hat
  binning in galaxy distance, using sub-samples taken from the Hubble
  Volume simulations. Using a simple model for the radial distribution
  expected for galaxies from a survey such as the Dark Energy Survey
  (DES), we show that this binning scheme will simplify analyses that
  will measure baryon acoustic oscillations within such galaxy
  samples. Comparing results from different binning schemes has the
  potential to provide measurements of the amplitude of the
  redshift-space distortions. Our analysis is relevant for other
  photometric redshift surveys, including those made by the Panoramic
  Survey Telescope \& Rapid Response System (Pan-Starrs) and the Large
  Synoptic Survey Telescope (LSST).
\end{abstract}


\section{Introduction} \label{sec:intro} 

The late-time acceleration of the expansion of the Universe has been
one of the most exciting cosmological discoveries in recent years
\citep{riess98,perlmutter99}. Understanding the nature of this
acceleration is one of the main challenges facing cosmologists. One of
the key observational methods that will be used to help meet this
challenge involves using Baryonic Acoustic Oscillations (BAO) in the
2-point galaxy clustering signal as a standard ruler to make precise
measurements of cosmological expansion. The acoustic signature has now
been convincingly detected \citep{percival01,cole05,eisenstein05}
using the 2dF Galaxy Redshift Survey (2dFGRS; \citealt{colless03}) and
the Sloan Digital Sky Survey (SDSS; \citealt{york00}). The detection
has subsequently been refined using more data and better techniques,
and is now producing interesting constraints on cosmological models
\citep{percival07a,percival07b,gaztanaga08,sanchez09,percival09}.

Some of the next generation of sky surveys, including the Dark Energy
Survey (DES {\tt www.darkenergysurvey.org}), the Panoramic Survey
Telescope and Rapid Response System (PanStarrs {\tt
  pan-starrs.ifa.hawaii.edu}), and the Large Synoptic Survey Telescope
(LSST {\tt www.lsst.org}), will use photometric techniques to estimate
galaxy redshifts, rather than more precise estimates from
spectroscopic emission lines. The larger uncertainties on galaxy
redshifts induce errors on inferred distances in the radial
direction. The amplitude of the power spectrum and correlation
function is reduced in the radial direction by this smoothing,
removing information. In this scenario, where little information
remains from fluctuations in the radial direction, it makes sense to
use the projected 2-pt functions in photometric-redshift slices as the
statistics to compare with models \citep{padmanabhan07,blake07}. The
projection does not completely remove problems caused by inferring
distances from velocity data (i.e. working in redshift-space).

The distribution of galaxies that we observe in sky surveys, where we
measure radial distances from spectroscopic or photometric redshifts, 
is not a true 3D picture. We observe an apparent clustering pattern in
{\it redshift-space}, which is systematically different from the true
distribution in {\it real-space} because redshifts of galaxies are
altered from their Hubble flow values by peculiar velocities. For
example, on large scales, the infall of galaxies onto collapsed
objects leads to an apparent enhancement of clustering in the radial
direction as galaxies are projected along their velocity vectors
\citep{kaiser87,hamilton98}. When we infer galaxy distances assuming
that the total velocity relative to the observer comes from the Hubble
expansion flow, the result is that we see a distorted (redshift-space)
density field.

For angular measurements, these redshift-space distortions can alter
the angular clustering in a redshift slice because the distortions are
correlated across the direction of projection. Although redshift-space
distortions are sub-dominant compared with photometric redshift
uncertainties, they give rise to a systematic effect, which needs to
be included when photometric redshift surveys are analysed
\citep{padmanabhan07,blake07}. This can complicate the analysis as the
size of the redshift-space distortions, and therefore of this effect,
is dependent on the cosmological model.  Consequently, for every model
to be tested against the data, we need to make a revised estimate of
the redshift-space effect.

\begin{figure}
\centering
\resizebox{0.9\columnwidth}{!}{\includegraphics{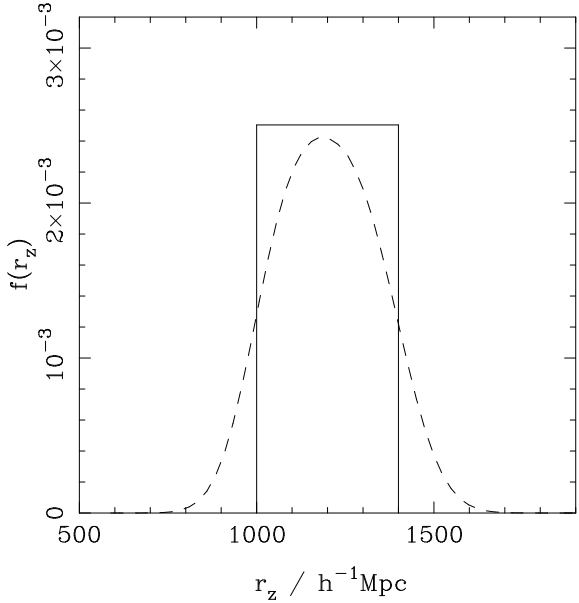}}
\caption{The radial distribution of galaxies selected in a bin of
  width $400\mpcoh$, calculated using photometric redshifts to
  estimate distances (solid line). This is compared against the
  distribution of true distances to these galaxies (dashed line)
  assuming a photometric redshift error of $\sigma_z=0.03(1+z)$. If
  the photometric redshifts of different galaxies are independent,
  then the expected projected correlation function of the photo-z
  selected sample, and a sample selected applying the dashed line as a
  selection function based on the true distances, are the
  same.\label{fig:photo-z-slice}}
\end{figure}
In this paper, we consider the simplified problem in the
plane-parallel approximation, and only consider linear redshift-space
distortions. Both photometric redshift errors and the random motion of
galaxies in clusters provide an additional convolution of the
overdensity field along the radial direction. While these effects need
to be corrected in any analysis, the required correction is easily
modelled and can be separated from the linear redshift-space
distortion effects. For a measurement of the projected clustering,
including such effects is equivalent to simply broadening the radial
window function with which the galaxies were selected. This is
demonstrated in Fig.~\ref{fig:photo-z-slice}. A top-hat bin in
photometric redshift gives the same expected projected correlation
function as simply applying the convolved version of the bin as a
selection function for the true distances. As we have to include a
window function anyway, we simply assume in this paper that this
window already includes the effects of both photometric redshift
errors and the random motion of galaxies in clusters. In the following
analysis, we therefore assume that there are no redshift errors
without loss of generality.

The layout of our paper is as follows. In Section~\ref{sec:2pt} we
analyse the projected overdensity field and redshift-space effects
upon it, both analytically (Section~\ref{sec:xi_p} \&~\ref{sec:pk_p})
and using Monte-Carlo simulations (Section~\ref{sec:mc_sim}). We then
consider how the recovered correlation function depends on galaxy
binning (Section~\ref{sec:dep_win}). Mock catalogues drawn from the
Hubble Volume simulation are constructed and analysed in
Section~\ref{sec:hv_sim} in order to validate this analytic work. We
incorporate hybrid selection functions based on both real and
redshift-space boundaries into the analysis in
Section~\ref{sec:hybrid}. In Section~\ref{sec:des} we consider a
non-uniform redshift distribution similar to that of future sky survey
DES, and the realistic implementation of our work is discussed in
Section~\ref{sec:discussion}.

\section{Projected 2-Point statistics of the overdensity field} 
\label{sec:2pt} 

\subsection{Correlation Function}  \label{sec:xi_p}

In order to simplify the problem, we assume that the clustering
strength does not change across the samples under consideration and
make the plane-parallel (distant observer) approximation, with
redshift-space distortions along the z-axis of a Cartesian basis.  In
the absence of redshift distortions the projected correlation function
is given by:
\begin{eqnarray}
  \xi_p(d_p) &=& 
    \langle\delta_p({\bf r_p})\delta_p({\bf r'_p})\rangle,\\
    &=& \int\int dr_zdr_z^{\prime}
    \phi(r_z)\phi(r_z^{\prime})\xi\left[d(r_z,r_z^{\prime},d_p)\right]
\label{eq:xi_proj_real}
\end{eqnarray}
where $d(r_z,r_z^{\prime},d_p) =
\sqrt{\left(r_z-r_z^{\prime}\right)^2+d_p^2}$, and subscripts $x$, $y$
and $z$ denote the direction along each Cartesian axis, and $p$
denotes projected quantities $p\equiv xy$. $\phi(r_z)$ is the radial
galaxy selection function, normalised such that
$\int\,dr_z\;\phi(r_z)=1$, and $\xi(d) = \langle\delta({\bf
  r})\delta({\bf r^{\prime}})\rangle$, where $\delta({\bf r})$ is
the overdensity of galaxies at real-space position ${\bf
  r}$. Throughout our paper we use ${\bf r}$ to describe a galaxy
position and ${\bf d}$ to describe the distance between two galaxies,
so, for example, $r_z$ is the position of a galaxy along the $z$-axis,
while $d_p$ is amplitude of the separation between two galaxies when
projected into the $x,y$-plane.

In reality, our radial position is determined via a redshift. In this
case, Eq.~(\ref{eq:xi_proj_real}) must be altered to
\begin{equation}
  \xi^s_p(d_p) = \langle\delta_p({\bf s_p})\delta_p({\bf s'_p})\rangle.
\label{eq:xi_proj_red1}
\end{equation}
The weighted, projected overdensity field $\delta_p(r_p)$ can now be
written
\begin{equation} \label{eq:del_proj_red} 
  1+\delta_p({\bf r_p}) =
    \int\,ds_z\;\phi(s_z)[1+\delta({\bf s})],
\end{equation}
where ${\bf s}=({\bf r_p},s_z)$ is the redshift-space position of each
galaxy and $\phi(s_z)$ gives the galaxy selection function along the
line of sight corresponding to $s_z$ (e.g. \citealt{peebles80}). 

\begin{figure}
\centering
\resizebox{0.9\columnwidth}{!}{\includegraphics{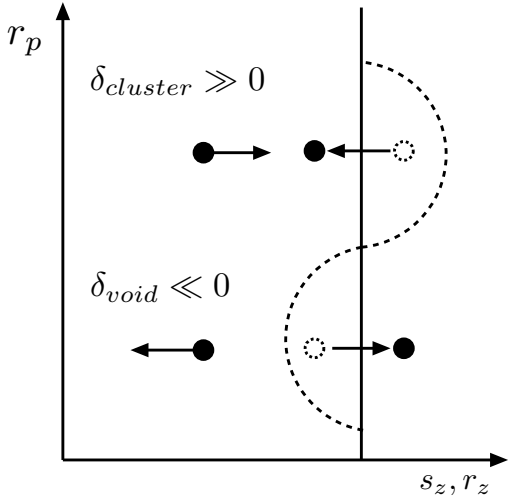}}
\caption{Schematic showing the boundary of a region selected in
  redshift-space (solid line) compared with the boundary of the same
  region in real-space (dashed line). The boundary is distorted in
  real-space around an overdensity and an underdensity. The positions
  of two galaxies whose apparent motion crosses the boundary are shown
  in redshift-space (solid circles) and in real-space (dashed
  circles). Note that, in this simplified picture where the under and
  overdensities have the same amplitude, the galaxy pair lost and the
  galaxy pair gained would contribute the same amount to the 3D
  real-space correlation function, following the dashed
  boundary. However, the projected clustering is different because we
  do not know the shape of the dashed line, and instead assume that
  the projection length is the same for all $r_x,r_y$. It is the 2D
  clustering strength of the boundary, and its correlation with the
  density field that is important, rather than the loss or gain of
  particular galaxy pairs. \label{fig:z-space}}
\end{figure}
The difference between the projection in redshift-space and real-space
is shown schematically in Fig.~\ref{fig:z-space}. An edge to a window
function (or a contour of constant galaxy density) that is straight in
redshift-space is systematically distorted in real-space. The edge of
the bin is itself clustered with a non-negligible projected
correlation function, i.e. the real-space boundary has a correlation
function that depends on ${\bf r_p}$. The inclusion or exclusion of
galaxies is balanced in terms of the 3D correlation function within
the boundary; while we lose voids, we gain clusters and these give the
same clustering signal. However, we assume that the projected field
has a constant projection length, and this implies that the
underdensity of the void will become larger (since we include less of
the galaxies) and the overdensity of the cluster becomes larger (since
we will include more of its galaxies). Thus the overall clustering
signal becomes stronger.

The apparent shift in galaxy positions caused by moving from real to
redshift space $(s_z-r_z)$ can be treated by Taylor expanding the
selection function \citep{fisher93}, which gives to first order
\begin{equation}  \label{eq:phi_red}
  \phi(s_z)=\phi(r_z) + \frac{d\phi(r_z)}{dr_z}(s_z-r_z).
\end{equation}
We consider this to be an Eulerian picture as it is based on apparent
galaxy motions. We can write
\begin{equation}  \label{eq:delp_proj_s_z}
  \delta_p({\bf r_p}) = \int\,dr_z\;
  \left[\phi(r_z)\delta({\bf r}) + 
    (s_z-r_z)\frac{\partial\phi(r_z)}{\partial r_z}\right]
\end{equation}
to first order in $\delta({\bf r})$.  Following linear theory,
$(s_z-r_z)$ can be written as a function of the overdensity field,
\begin{equation}  \label{eq:smr}
  (s_z-r_z)=-\beta\frac{\partial}{\partial r_z}\nabla^{-2}\delta({\bf r}),
\end{equation}
where $\beta\equiv f/b$, with $f$ being the logarithmic derivative of
the linear growth rate with respect to the logarithm of the scale
factor, and $b$ the galaxy bias. We therefore have that
\begin{equation}  \label{eq:delp_proj_delta}
  \delta_p({\bf r_p}) = \int\,dr_z\;
  \left[\phi(r_z) - \beta\frac{\partial\phi(r_z)}{\partial r_z}
    \frac{\partial}{\partial r_z}\nabla^{-2}\right]\delta({\bf r}).
\end{equation}
If we think of $\phi(s_z)$ as setting up boundaries in $s_z$, then
substituting Eq.~(\ref{eq:delp_proj_delta}) into
Eq.~(\ref{eq:xi_proj_red1}) shows that we can expect coherent apparent
galaxy motion across these boundaries. Correlations between galaxies
moved into the sample by the redshift-space distortions, and those
already within the sample, give rise to cross terms from the two terms
in Eq.~(\ref{eq:delp_proj_delta}). The second term in
Eq.~(\ref{eq:delp_proj_delta}) also adds a component to the projected
correlation function from the coherence of the velocities at different
points on the boundary. We see that, even with constant $\phi(s_z)$
within a fixed interval, redshift-space distortions can still affect
the correlation function of the volume within the sample due to the
motion of galaxies across the boundary. Modelling the effect of
redshift-space distortions based on predicting galaxy motions
(e.g. \citealt{regos95}) is difficult because we need to correlate
multiple points on the boundary and internal locations within the bin.

In addition to the Eulerian picture given by
Eq.~(\ref{eq:delp_proj_delta}), we can also consider a Lagrangian
picture based on the redshift-space overdensity field that we wish to
project. Following this equivalent picture, we can work directly with
redshift-space overdensities using Eq.~(\ref{eq:xi_proj_red1}),
\begin{equation}
  \xi^s_p(d_p) =  \int\int ds_zds_z^{\prime}
    \phi(s_z)\phi(s_z^{\prime})\xi^s\left[d(s_z,s_z^{\prime},d_p)\right].
  \label{eq:xi_proj_red}
\end{equation}
In the plane-parallel approximation, we can use the redshift-space
correlation function of equation 5 of \citet{hamilton92} as input into
the projection equation.
\begin{equation}  \label{eq:xi_mu}
  \xi^s({\bf d}) = \xi_0(d)P_0(\mu)+\xi_2(d)P_2(\mu)+\xi_4(d)P_4(\mu),
\label{eq:xi_s}
\end{equation}
where
\begin{eqnarray}
  \xi_0(d) &=& (b^2+\frac{2}{3}bf+\frac{1}{5}f^2)\xi(d), \\
  \xi_2(d) &=& (\frac{4}{3}bf+\frac{4}{7}f^2)[\xi(d)-\xi'(d)], \\
  \xi_4(d) &=& \frac{8}{35}f^2[\xi(d)+\frac{5}{2}\xi'(d)-\frac{7}{2}\xi''(d)],
\end{eqnarray}
$P_i$ are the standard Legendre polynomials, and
\begin{eqnarray}
  \xi'(d)&\equiv&3d^{-3}\int^d_0\xi(d')(d')^2dd',\\
  \xi''(d)&\equiv&5d^{-5}\int^d_0\xi(d')(d')^4dd'.
\end{eqnarray}
$b$ is the large-scale bias of the galaxy population being considered,
$f$ is the standard dimensionless linear growth rate, $\xi$ is the
3-dimensional real-space correlation function, and $\mu$ is the cosine
of the angle between the separation along the line of sight and the
transverse separation, $\mu\equiv|s_z-s_z^{\prime}|/d$. One strong
advantage of the Lagrangian framework is that it is straightforward to
determine the projected correlation function, even when the galaxy
selection function is discontinuous. This allows simple comparison
between the results one expects to obtain with and without
redshift-space distortions.

\subsection{The Limber approximation}

For pairs of galaxies, we can define the mean
$m_z\equiv(r_z+r_z^{\prime})/2$ and separation along the z-axis
$d_z\equiv r_z-r_z'$. For a survey whose depth is larger than the
correlation length, and with a slowly varying selection function, so
that $\phi(r_z)\simeq\phi(r_z')\simeq\phi(m_z)$,
Eq.~(\ref{eq:xi_proj_real}) reduces to the Limber equation in
real-space ($s_z-r_z=0$)
\begin{equation} \label{eq:limber1}
  \xi_p(d_p) = \int_{-\infty}^{+\infty}dm_z\; \phi^2(m_z) 
  \int_{-\infty}^{+\infty}dd_z\;\xi\left(\sqrt{ d_p^2 + d_z^2}\right).
\end{equation}
We see that, for the Limber approximation, $\phi$ is a function of
$m_z$ alone, and the integrals over $dm_z$ and $dd_z$ in
Eq.~(\ref{eq:limber2}) are separable. In redshift-space, a similar
reduction of Eq.~(\ref{eq:xi_proj_red}) gives
\begin{displaymath}
  \xi^s_p(d_p) = 
    \int_{-\infty}^{+\infty}dm_z \int_{-\infty}^{+\infty}dd_z 
    \left[\phi(m_z) - \beta\frac{\partial\phi(m_z)}{\partial m_z}
    \frac{\partial}{\partial r_z}\nabla^{-2}\right]^2 
\end{displaymath}
\begin{equation} \label{eq:limber2}
  \hspace{20mm}\times
    \;\xi\left(\sqrt{d_p^2+d_z^2}\right),
\end{equation}
if we expand redshift-space distortions in $(s_z-r_z)$, or 
\begin{equation} \label{eq:limber3}
  \xi^s_p(d_p) = \int_{-\infty}^{+\infty}dm_z\; \phi^2(m_z) 
  \int_{-\infty}^{+\infty}dd_z\;\xi^s\left(\sqrt{ d_p^2 + d_z^2}\right),
\end{equation}
in the Lagrangian picture. Because no galaxies are lost or gained
moving from real-space to redshift-space, the result of the integral
over $d_z$ is the same in real or redshift space, so we see that in
this approximation there are no redshift-space effects. But, as we
show later, this picture is too simplistic to be applied to the
analysis of future data sets.

\subsection{Power Spectra}  \label{sec:pk_p}

In \citet{padmanabhan07}, the projection of the 2-pt clustering was
analysed through the power spectrum. We now consider such an approach
in the plane-parallel approximation and for a Cartesian basis. Taking
the Fourier transform of $\delta({\bf s})$ in
Eq.~(\ref{eq:del_proj_red}) gives
\begin{equation} \label{eq:del_proj} 
  \delta_p({\bf r_p}) =
    \int\,ds_z\;\phi(s_z) \int\frac{d^3k}{(2\pi)^3}
    \delta({\bf k})e^{-i{\bf k}\cdot{\bf s}}.
\end{equation}
We now define a window function
\begin{equation}
  W(k_z)=\int\,ds_z\;\phi(s_z)e^{-ik_zs_z},
\end{equation}
and use statistical isotropy and homogeneity within the definition of
the power spectrum $\langle\hat{\delta}({\bf k})\hat{\delta}^*({\bf
  k'})\rangle=P(k)\delta_D({\bf k}-{\bf k'})$, where $\delta_D$ is the
Dirac delta function. We assume that the power spectrum does not
evolve over the volume covered by the window\footnote{This is true if
  analysing a single time slice from a simulation}. Taking the 2-point
function of the projected overdensity (Eq.~\ref{eq:del_proj}) gives
\begin{eqnarray}
  \xi_p(d_p) & = & 
    \langle \hat{\delta}_p({\bf r_p}) \hat{\delta}({\bf r_p^{\prime}}) \rangle \\
    & = & \int \frac{dk^3}{(2\pi)^3} W^2(k_z) P({\bf k})  \label{eq:r-2pt-proj1}
    e^{-i{\bf k_p}\cdot({\bf r_p-r_p^{\prime}})}.
\end{eqnarray}
The projected overdensity can be written in terms of a 2D power
spectrum $P_p(k_p)$,
\begin{equation}  \label{eq:r-2pt-proj2}
  \xi_p(d_p) = 
    \int\frac{dk_x\,dk_y}{(2\pi)^2}P_p(k_p)
    e^{-i{\bf k_p}\cdot({\bf r_p-r_p^{\prime}})}.
\end{equation}
If we compare Eqns.~(\ref{eq:r-2pt-proj1}) \&~(\ref{eq:r-2pt-proj2}),
we see that
\begin{equation} \label{eq:Pkxy}
  P_p(k_p)=\int \frac{dk_z}{(2\pi)}W(k_z)^2P\left(\sqrt{k_p^2+k_z^2}\right).
\end{equation}
Note that the power $P({\bf k})$ depends on the amplitude of the full
3-dimensional wavevector, and so is dependent on $k_p$.

Using Eq.~(\ref{eq:phi_red}) to include redshift-space distortions,
the window $W(k_z)$ has an extra term, 
\begin{equation}
  W(k_z)=\int\,dr_z\;
    \left[\phi(r_z)+(s_z-r_z)\frac{d\phi(r_z)}{dr_z}\right]
    e^{-ik_{z}r_{z}}.
\end{equation}
In Fourier space, $(s_z-r_z)=-\beta(k_z^2/k^2)\delta({\bf r})$, so we
can expand $\delta({\bf s})$ to 1st order in $\delta({\bf r})$,
leaving a new window function for Eq.~(\ref{eq:Pkxy})
\begin{equation}
  W(k_z)=\int\,dr_z\;
    \left[\phi(r_z)-\beta\left(\frac{k_z}{k}\right)^2
    \frac{d\phi(r_z)}{dr_z}\right]e^{-ik_{z}r_{z}}.
\end{equation}

If we drop the plane-parallel approximation and expand in Spherical
Harmonics, the standard result \citep{peebles73} is
\begin{equation}
  \langle |a_{lm}|^2 \rangle =
    \frac{1}{2\pi^2}\int\,dk\;k^2P(k)W^2(k),
\end{equation}
where
\begin{equation}
  W(k)=\int\,dr\;
  \phi(r)j_l(kr)+\frac{\beta}{k}\frac{d\phi(r)}{dr}j_l'(kr).
\end{equation}
Here the $l$ dependence is contained within $W(k)$, while in
Eq.~(\ref{eq:Pkxy}), it was the power that depended on
$k_p$. Eq.~(\ref{eq:Pkxy}) could have been rewritten by changing the
variable of the convolution integral to $k$ to match.

\subsection{Monte-Carlo simulations of the projection effect}  
\label{sec:mc_sim}

In order to test the projection formulae presented in
Sections~\ref{sec:xi_p} \&~\ref{sec:pk_p} without redshift-space
distortions, we have used Monte-Carlo realisations of
$\delta$-function real-space correlation functions in a similar vein
to that of \citet{simpson09}. We work in a plane parallel
approximation throughout and construct a real-space 3D
$\delta$-function correlation function at an arbitrary location $d_0$
such that
\begin{equation}\label{eq:xi-delta}
  \xi(d)=\delta_D(d-d_0)\; \xi_0,
\end{equation}
where $\delta_D$ is the standard Dirac delta function. We do this by
introducing a pre-determined excess of data pairs at the location
$d_0$.  The number of excess pairs we introduce depends on the value
of $\xi(d_0)$ we require and is determined using the natural estimator
$\xi=D/R-1$. For example, if we have a uniform distribution of data
and random pairs with $100,000$ pairs per bin of separation, we would
require an excess of $10,000$ data pairs at the location $d_0$ for
$\xi(d_0)=0.1$. In doing this we create an unnormalised 3D
$\delta$-function correlation function.

Changing the variables in the inner integral of Eq.~(\ref{eq:limber1})
to be a function of 3D pair separation $d$ gives 
\begin{equation} \label{eq:limber4}
  \xi_p(d_p) = \int \int_V \,dm_z \,dd\;\;
  \phi^2(m_z)\; \frac{2\xi(d)\,d}{\sqrt{d^2-d_p^2}},
\end{equation}
and is simplified for the $\delta$-function case such that
\begin{equation} \label{eq:limber-delta}
  \xi_p(d_p) = \frac{2}{\pi d_0}\int \,dm_z\; \phi^2(m_z) \, 
  \xi_0 \frac{d_0}{\sqrt{d_0^2-d_p^2}}.
\end{equation}
The factor $1/\pi d_0$ accounts for the fact that the
$\delta$-function real-space correlation function was unnormalised.
By introducing a radial window, we are preferentially selecting pairs
of galaxies from the sample. A further volume reduction normalisation
is required in Eq.~(\ref{eq:limber4}) to account for this. The excess
probability of finding two galaxies in areas $\delta A_1$ and $\delta
A_2$ with a 2D projected separation $d_p$ is the sum of all the
probabilities of finding two galaxies in volumes $\delta V_i$ and
$\delta V_j$ along the radial axis at {\it all} 3D separations
$d$. That is,
\begin{equation} \label{eq:vol-norm}
  1+\xi_p(d_p) = \frac{\bar{n_V}^2}{\bar{n_A}^2} \frac{1}{\delta A_1 \delta A_2} 
    \left ( \sum_i \sum_j [1+\xi(d_{ij})] \delta V_i \delta V_j \right ).
\end{equation}

\begin{figure}
\centering
\includegraphics[width=0.9\columnwidth]{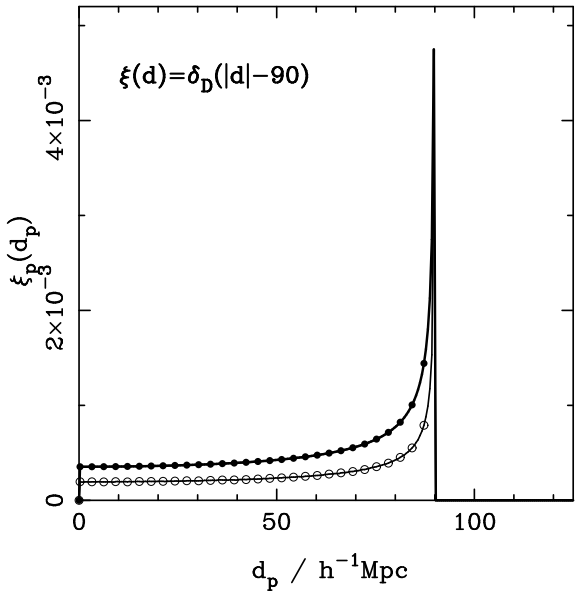}
\caption{Projected correlation functions calculated for a
  3-dimensional $\delta$-function correlation function centred on
  $d_{3D}=90\mpcoh$, with no radial window (solid symbols) and with a
  top-hat window in radial distribution of width $100\mpcoh$ (open
  symbols). Models calculated using Eq.~(\ref{eq:limber-delta}) are
  shown by the solid lines.\label{fig:xi_df_slice}}
\end{figure}
In Fig.~\ref{fig:xi_df_slice} we show the clustering expected for a
projection of a density field created from a $\delta_D$-function 3D
correlation function in the case where there is no window function
(solid symbols) and for a window function of width 100$\mpcoh$
(open symbols). The excess of pairs that exists at a single scale
in 3D is projected onto a range of scales, up to and including this scale, 
in 2D. The projection window leads to a damping of power on all scales. 
This effect depends upon the window size; as we move to smaller projection 
windows the effect of the projection is decreased and 
$\xi_{2D} \rightarrow \xi_{3D}$. The projection of a more
general density field, where there is clustering on a range of scales,
can be considered as the linear combination of the projections of a
series of $\delta_D$-function 3D correlation functions. The trends
observed in this analysis will help us to interpret the behaviour of 
the projected correlation function in the more general situation analysed
in later sections.

\section{Binning galaxy samples} \label{sec:dep_win}

Future surveys will automatically have a standard selection function
caused by the changing cosmological volume, the number density of
galaxies as a function of redshift, and selection effects such as a
magnitude limit below which we cannot observe galaxies or obtain
accurate photometric redshifts. In addition to this distribution we
will wish to bin galaxies based on their photometric redshifts in
order to analyse the evolution of galaxy properties and/or cosmology
across the sample. We now consider how the way in which this
sub-division is applied affects the importance of redshift-space
distortions.

One simple approach would be to bin galaxy positions in redshift,
equivalent to a {\bf top-hat binning}. Such galaxy selection means
that galaxy pairs, where galaxies lie in different bins, are not
included in the estimate of the correlation function. This exclusion
of pairs leads to the observed difference between the projected
real-space and redshift-space correlation function, as described in
Section~\ref{sec:xi_p}. An alternative to this approach, considered
here, would be to bin galaxy pairs rather than individual galaxies.

\begin{figure}
\centering
\resizebox{0.65\columnwidth}{!}{\includegraphics{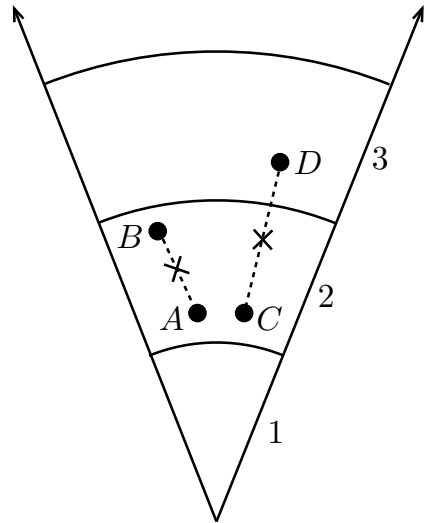}}
\caption{A schematic representation showing how galaxy pairs are
  selected using top-hat and pair-centre binning schemes. Using a
  top-hat binning scheme, where galaxy pairs are selected according to
  the position of each individual galaxy, pair $AB$ would be placed in
  redshift bin 2, whereas pair $CD$ would not be placed in any bin,
  and would simply not be counted in an analysis. In contrast, the
  pair-centre binning scheme would place both pairs in bin
  2. \label{fig:pc_bin_schematic}}
\end{figure}

A simple argument shows that in an ideal situation, applying a binning
based on the centre of galaxy pairs in the radial direction, which
hereafter we refer to as {\bf pair-centre binning}, can completely
remove the effect of redshift-space distortions while retaining
information about the evolution of the correlation function. A
schematic representation of this binning scheme is shown in
Fig.~\ref{fig:pc_bin_schematic}.  Consider the galaxy pair defined by
galaxies $A$ and $B$: the positions of both galaxies and their
pair-centre are within redshift slice 2. This pair would therefore be
included in analyses conducted on this slice in both top-hat and
pair-centre binning schemes. The positions of galaxies $C$ and $D$
span two separate redshift slices and therefore the pair they define
would not be included in an analysis of either slice 2 or 3 when using
a top-hat binning scheme. However, this pair would be included in an
analysis of slice 2 when using the pair-centre binning scheme. This
schematic demonstrates both the pair-centre binning scheme and the
fact that such a scheme includes all pairs within an analysis.

Suppose that we have a clustered distribution of $D$ galaxy pairs of
separation $r$ with a uniform sampling function along the z-axis in a
large volume that would contain $R$ pairs if galaxies were randomly
distributed. Because of the large volume assumption, we can assume
that boundary effects for this sample are negligible. Therefore,
redshift-space distortions have no effect for the full catalogue for
which our estimate of $\xi_p(d_p)$ is
$\hat{\xi_p}(d_p)=D/R-1$. Now suppose the sample is split into $n$
sub-samples, based on the redshift-space positions of the centres of
the pairs within equal volumes, chosen independently of the observed
galaxy distribution. Then all pairs are still counted in some bin;
none are lost or gained as opposed to galaxy based selection
functions. For the subsamples, $\langle D'\rangle=D/n$, $R'=R/n$, and
$\langle\xi_p(d_p)\rangle$ is unchanged from the value for the full
sample. This is true regardless of bin size. The key
difference here, compared with considering a set of bins based on
galaxy selection, is that no pairs are left out, so the expected
correlation function has to be the same for all bins.

For a sample where we do not know the true distance to each galaxy, but
instead rely on photometric redshifts, binning based on apparent pair
centre will also remove redshift-space distortions. The above argument
based on pair conservation will also hold in this
situation.

\begin{figure}
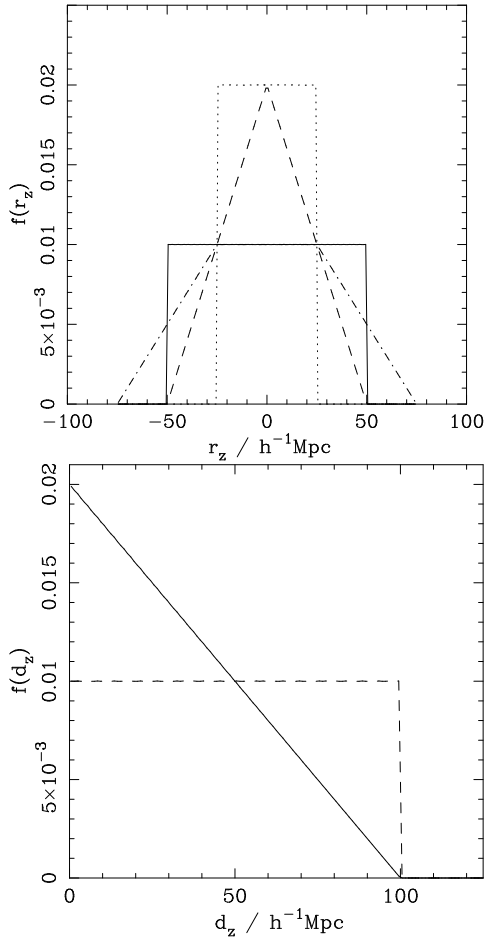

\centering
\resizebox{0.75\columnwidth}{!}{\includegraphics{bin_dist_r.ps}}
\resizebox{0.75\columnwidth}{!}{\includegraphics{bin_dist_d.ps}}
\caption{Top panel: the normalised radial distribution of galaxies
  (solid line) and pair-centres (dashed line) for the distribution of
  galaxies in a top-hat bin of width $100\mpcoh$. These are compared
  with the distributions of galaxies (dot-dash line) and pair-centres
  (dotted line) for galaxies whose pair-centre is within a $50\mpcoh$
  bin, and with $d_z<100\mpcoh$. Bottom panel: comparison of the
  radial pair separations ($d_z$), between top-hat (solid line) and
  pair-centre (dashed line) binning. \label{fig:bin_dist}}
\end{figure}
The radial distributions of galaxies and pair-centres along with the
distribution of radial pair separations for top-hat and pair-centre
binning schemes are compared in Fig.~\ref{fig:bin_dist}. For measuring
the radial evolution of clustering through binned projected
correlation function measurements, there is no obvious advantage to
either scheme. This is particularly true when photometric redshifts
are used to estimate radial positions, as it is then impossible to
select galaxies from non-overlapping radial bins (see
Section~\ref{sec:discussion}).

We therefore see that we can add boundaries based on pair-centres and
analyse projected clustering in bins without being affected by
redshift-space distortions. However, there are two problems with
applying this approach in practise:
\begin{enumerate}
\item galaxy pairs of wide separation now have to be included,
\item galaxy surveys typically have flux limited boundaries, which
  cause redshift dependent effects that cannot be removed by any
  binning. However, this effect can be removed by $k$-correcting the
  observed luminosities and cutting the sample at a more stringent
  $k$-corrected luminosity limit. We now investigate this further.
\end{enumerate}

\subsection{Flux-limited Selection Functions}
\label{sec:mag}

Peculiar velocities can directly influence galaxy brightness through
relativistic beaming, but such effects are small for typical galaxy
peculiar velocities. Redshift distortions would additionally change
the apparent magnitudes through the $k$-correction, potentially
causing galaxies to either enter or exit flux-limited samples. The
change in apparent magnitude will correlate with bulk-flow motions and
thus the boundary of the survey in real-space will fluctuate in a
manner analogous to that described in Fig.~\ref{fig:z-space}. In this
situation, the amplitude of the effect and whether it enhances or
reduces the real-space clustering signal will depend on galaxy type
and the band used for detection, but for a homogeneous sample of
galaxies (e.g. Luminous Red Galaxies) one would expect that this
effect will be significant.

This redshift-space effect is simple to remove - $k$-corrections
derived by fitting to galaxy spectra will correct for spectral shifts
caused by both the Hubble flow and any peculiar velocities. It
therefore makes sense to select galaxy samples after applying the
$k$-correction, and cutting back from survey boundaries based on
apparent magnitude, until no galaxies outside the original sample
would be expected to pass the revised boundary. This is not as onerous
as it sounds as one has to do this to create true volume-limited
catalogues.

Given purely photometric data, $k$-corrections can only be estimated
given a photometric redshift and spectral-type fit, and are therefore
unreliable for individual galaxies.  For this reason, and the fact
that cutting back from the survey boundary removes a large amount of
data, $k$-corrections have not always been applied to apparent
magnitude limits (e.g. \citealt{R09} select galaxies with de-reddened
$r < 21$ for their parent sample).  We therefore now consider the
amplitude of the effect. One can express the fluctuation in magnitude,
$\delta m$, as
\begin{equation}
  \delta m = {\rm d }k_{corr}/{\rm d}z \delta z
\end{equation}
where $\delta z$ is the magnitude of the redshift distortion.  This
will cause fluctuations in the effective depth of the survey such that
\begin{equation}
  DM(z_{eff}) - DM(z) = \delta_m
\end{equation}
where $DM(z)$ is the distance modulus, $z_{eff}$ is the effective
depth and $z$ would be the predicted depth.  The SDSS DR7 photometric
redshift table includes estimated $r$-band $k$-corrections for
every galaxy.  Studying galaxies with type-value equal to 0 (the most
early-type), one can determine that ${\rm d }k_{corr}/{\rm d}z \sim
3.3$ at $z = 0.4$.  For an arbitrary $\delta z$, this ${\rm d
}k_{corr}/{\rm d}z$ yields $z_{eff} - z = 0.5\delta z$.  For example,
assuming bulk flows have a velocity $\sim 10^{3}$km/s --- thereby
imparting redshift distortions at the $\sim1\%$ level ($\delta z =
0.004$) --- they impart coherent fluctuations in apparent magnitude
equivalent to 0.013 magnitudes (in the $r$-band).  At $z = 0.4$, these
fluctuations in magnitude imply a change in the survey depth of
$z_{eff} -z = 0.002$ (0.5$\%$).  Thus, the redshift distortions caused
by selecting a flux-limited sample of galaxies can be as large as
50$\%$ of those caused by selecting a sample in redshift.  Therefore,
even for a flux limited selection function, redshift distortions may
be important.  The size of the effect depends on the slope of
$k_{corr} (z)$, and one can minimise the effect by carefully choosing
the band used for selection and the type(s) of galaxies included in
the sample.  (One can envision cases where slope of the average
$k$-correction is zero, thus removing any effect even before applying
$k$-corrections.)

\begin{figure*}
\centering
\resizebox{0.75\textwidth}{!}{\includegraphics{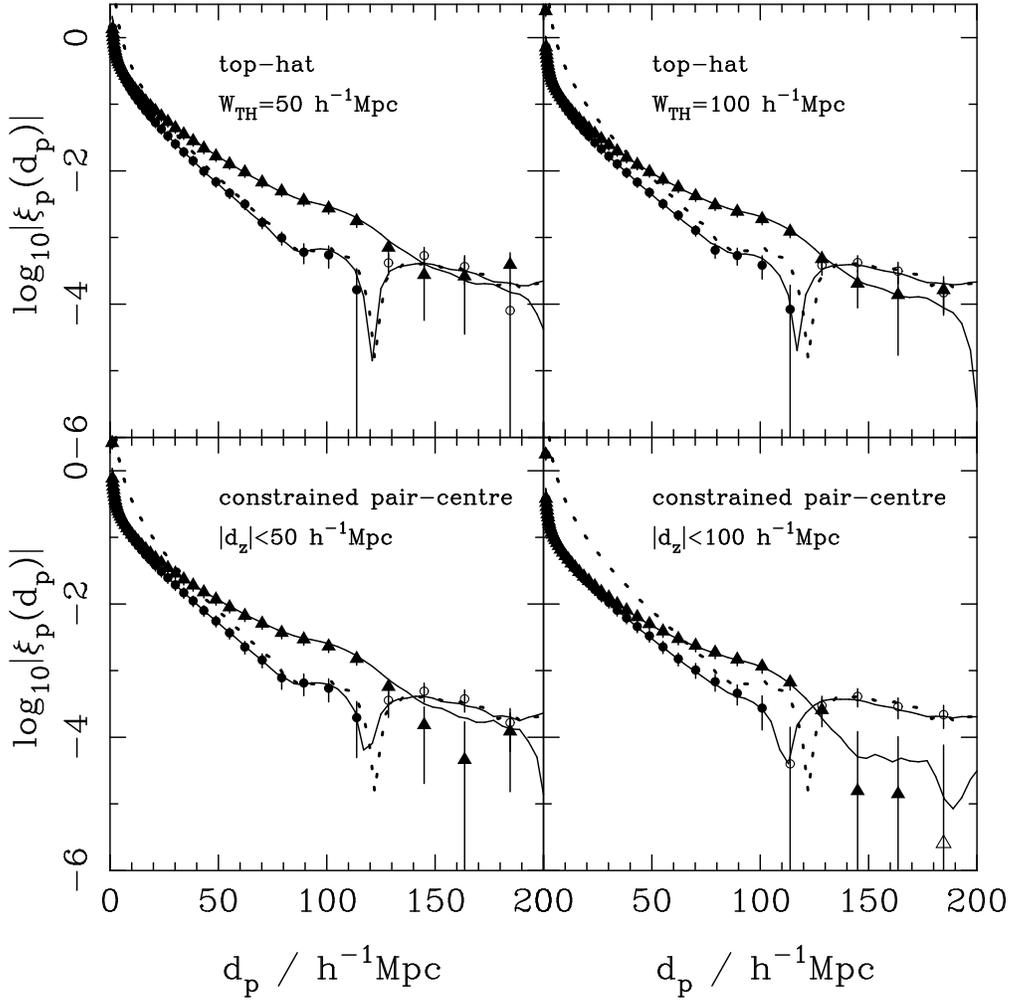}}
\caption{Top row: Correlation functions calculated from Hubble Volume
  data in radial bins of width 50$\mpcoh$ and 100$\mpcoh$. Solid
  symbols are plotted where the correlation function is positive,
  while open symbols show where the correlation function is
  negative. Bottom row: Correlation functions calculated from Hubble
  Volume data for galaxy pairs selected based on constrained
  pair-centre binning scheme with both their pair-centres and radial
  separation less than 50$\mpcoh$ or 100$\mpcoh$. $1\sigma$ error bars
  are plotted in both cases, assuming that the slices analysed draw
  correlation functions from a Gaussian distribution. The dotted line
  gives the 3D HV correlation function, $\xi_{HV}$, (plotted assuming
  $r=r_p$) as measured from the simulation.  The solid lines denote
  the projected correlation function one expects in real-space with
  $\xi_{HV}$ as the 3D correlation function (lower curves in each
  panel), and in redshift-space using Eqns.~(\ref{eq:xi_proj_red})
  \&~(\ref{eq:xi_mu}) to estimate the 3D redshift-space correlation
  function from $\xi_{HV}$ (upper curves in each
  panel). \label{fig:xi_hv_slice}}
\end{figure*}

\section{Analysis of Hubble Volume simulations}  \label{sec:hv_sim}

In order to test the effect of redshift-space distortions on the
projected correlation function for a realistic non-linear distribution
of galaxies, we have analysed results from the $\Lambda$CDM Hubble
Volume (HV) simulations \citep{evrard02}. The $\Lambda$CDM HV
simulation, covering a $(3000\mpcoh)^3$ box, assumes a cosmological
model with $\Omega_m=0.3$, $\Omega_{CDM}=0.25$, $\Omega_b=0.05$,
$\Omega_{\Lambda}=0.7$, $h=70$, $\sigma_8=0.9$, \& $n_s=1$.

We make a number of simplifications in order to help with the
calculation of projected real-space and redshift-space correlation
functions. For each sample to be analysed, along the two
non-projection axes, we use the periodic nature of the numerical
simulation to eliminate boundaries. This means that we can confidently
use the natural estimator $\xi+1=D/R$, where the expected number of
galaxy pairs in the absence of clustering $R$ can be calculated
analytically. We also do not introduce a galaxy-bias model, and assume
that galaxies Poisson sample the matter particles. The inclusion of
such a model would not alter the conclusions of this work.

We start by applying a top-hat selection function to the galaxy
positions, calculating projected correlation functions for window
widths 50$\mpcoh$ and 100$\mpcoh$ in real and redshift
space. Fig.~\ref{fig:xi_hv_slice} shows the correlation function after
reducing noise by averaging over $30$ samples (100$\mpcoh$ bins) or
$60$ samples (50$\mpcoh$) bins. In real-space the projected
correlation function tends towards the 3D correlation function at
large scales, as expected. In line with the analysis presented in
Section~\ref{sec:mc_sim}, the scale at which $\xi_p$ becomes $\sim
\xi_{3D}$ is larger for the 100$\mpcoh$ bin.  For each bin size, the
inclusion of redshift-space distortions clearly has a strong effect
and this effect grows dramatically as the scale gets larger.  Notably,
it is larger even than the effect of redshift-space distortions on the
3D spherically averaged correlation function (or power spectrum). The
effect is enhanced in the narrower projection window. As well as
increasing the amplitude of the projected correlation function, we see
that redshift-space distortions also act to wash out the baryon
acoustic oscillation signal.

Selecting galaxy pairs solely based on the position of their
pair-centre removes the effect of redshift-space distortions. To see
this, suppose we split along the projection axis into $N$ slices, and
average the $DD$ counts over all slices. Then the average is
independent of $N$ as all pairs are counted however many bins are
selected. In addition, the periodic nature of the simulation means
that no pairs are gained or lost between real-space and
redshift-space: we always count all pairs of galaxies, so there will
be no change in the measured correlation function. As explained in
Section~\ref{sec:mag}, if we select based on an apparent magnitude
limit, we can remove redshift distortions by applying a more stringent
magnitude limit based on $k$-corrected luminosities. Here we have to
cut the luminosity limit back to make sure that the new sample is
complete, in that it contains all of the possible galaxies. However,
there is a further practical problem in that including galaxy pairs
with wide radial separation might complicate the modelling of
cosmological evolution required to fit the correlation
function. Consequently, it might be difficult to analyse the measured
correlations function for a pair-centre binned sample in practice.
 
We therefore introduce a {\bf constrained pair-centre} binning scheme
that includes an upper limit on the pair separation along the
projection axis, in addition to pair-centre binning. This is
equivalent to locating {\em each} galaxy included in the analysis in
the centre of a top-hat bin. We should expect that the effect of
redshift-space distortions will be reduced compared with binning
galaxy distributions in a top-hat with the same width, as boundaries
will only affect galaxy pairs with the maximum radial separation,
whereas for top-hat bins they affect galaxy pairs with a range of
radial separations (see Fig.~\ref{fig:bin_dist}). Results calculated
using this binning scheme are shown in
Fig.~\ref{fig:xi_hv_slice}. Here we see that the effect of
redshift-space distortions is reduced, especially for the larger
$|d_z|$ limit.

\begin{figure}
\centering
\resizebox{0.9\columnwidth}{!}{\includegraphics{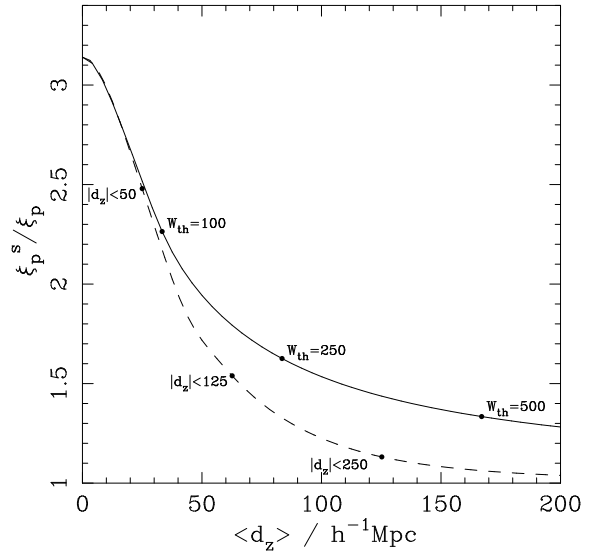}}
\caption{The expected ratio of the projected correlation functions in
  redshift-space and in real-space, averaged for ``angular''
  separations between $40\mpcoh$ and $80\mpcoh$, as a function of bin
  width. The solid line show the difference as a function of the width
  of the top-hat window. The dashed line show the result for
  constrained pair-centre binning as a function of an additional
  constraint placed on the radial galaxy separation. We have plotted
  results (and therefore matched filters) as a function of the mean
  radial galaxy separation. \label{fig:xis_cmpr_bin}}
\end{figure}
In order to investigate the effect of different binning schemes
further, Fig.~\ref{fig:xis_cmpr_bin} shows a comparison on the
large-scale redshift-space and real-space correlation function
amplitude. These are averaged for galaxy separations between
$40\mpcoh$ and $80\mpcoh$. We have plotted these as a function of
average radial galaxy separation, in order to compare filters in an
unbiased way. We clearly see that, when binning radially using the
constrained pair-centre binning scheme, the effect of redshift-space
distortions is significantly reduced.

\begin{figure}
\centering
\resizebox{0.9\columnwidth}{!}{\includegraphics{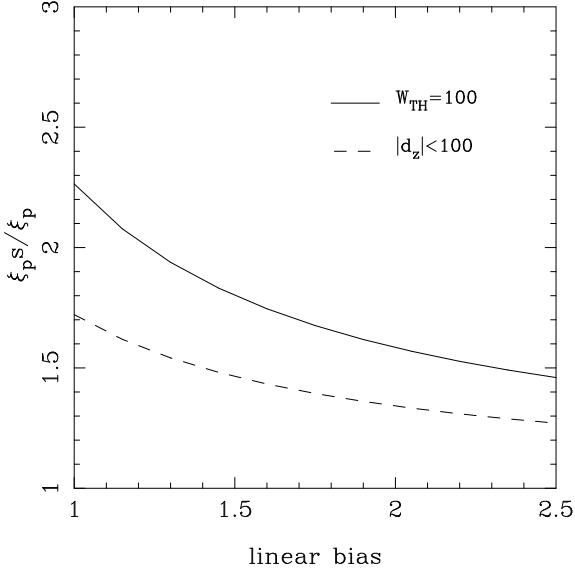}}
\caption{The expected ratio of the projected correlation functions in
  redshift-space and in real-space, averaged for ``angular''
  separations between $40\mpcoh$ and $80\mpcoh$, as a function of
  galaxy bias, assuming a $\Lambda$CDM cosmology with
  $\Omega_m=0.25$. The solid line show the difference as a function of
  the width of the top-hat window. The dashed line show the result for
  constrained pair-centre binning as a function of the
  additional constraint placed on the radial galaxy separation. As in
  Fig.~\ref{fig:xis_cmpr_bin}, we have plotted results (and therefore
  matched filters) as a function of the mean radial galaxy
  separation. \label{fig:xis_cmpr_bias}}
\end{figure}
The relative importance of redshift-space distortions depends on the
average galaxy bias of the populations being considered; there is a
balance between the impacts of $b$ and $f$ in Eq.~(\ref{eq:xi_mu}). In
order to demonstrate this, Fig.~\ref{fig:xis_cmpr_bias} shows that the
relative effect of redshift-space distortions decreases as the bias of
the galaxy sample analysed increases. This explains why the effect of
redshift-space distortions was reduced in the work of
\cite{baldauf09}.

In this section, we have considered the cases of a top-hat or
pair-centre galaxy selection. We have argued that while, in principle,
pair-centre binning removes the effects of redshift-space distortions
provided $k$-corrections are included when magnitude limits are
applied, there are good reasons to remove galaxies of wide separation
if we are to measure the evolution in the correlation
function. Therefore, in Section~\ref{sec:des}, we will test how these
binning schemes work when the background galaxy distribution has a
more realistic radial distribution, similar to that expected for a
survey like the Dark Energy Survey. Before we can do that, we need to
consider the case where we have a boundary that consists of a mix of
real- and redshift-space constraints.

\section{Dealing with Hybrid Selection Functions}
\label{sec:hybrid}

In practice, the radial selection function will be dependent on both
observational constraints such as the limiting apparent magnitude of
the survey, and additional binning. One expects that the boundary
based on observational constraints can be treated as a real-space
boundary (though, this even, is not so simple; see Section~\ref{sec:mag}).
Thus, when one applies a top-hat selection in redshift to an observed
sample of galaxies, the resulting boundaries of the selection function
will include both real-space and redshift-space components.

\begin{figure}
\centering
\resizebox{0.9\columnwidth}{!}{\includegraphics{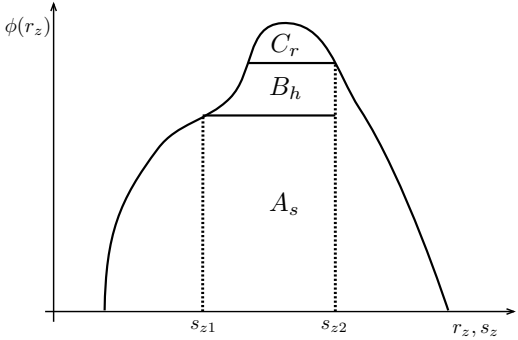}}
\caption{Schematic representation of an evolving real-space radial selection 
         function with populations $A_s$, $B_h$ and $C_r$ defined according to 
         where a top-hat bin with redshift-space boundaries at $s_{z1}$ and 
         $s_{z2}$ intersect the radial selection. Populations have boundaries
         in: $A_s$ redshift-space, $B_h$ hybrid-space and $C_r$ real-space.
         \label{fig:sch_phi_real_red}}
\end{figure}         
Fig.~\ref{fig:sch_phi_real_red} shows a schematic representation of a
top-hat selection in redshift made at positions $s_{z1}$ and $s_{z2}$
along a non-uniform real-space radial selection function.  It shows
that we can split galaxies within this bin into three sub-samples,
with different boundaries:
\begin{itemize}
  \item $A_s$ (redshift-redshift): Selected with both boundaries in redshift-space.
  \item $B_h$ (redshift-real): Selected with one boundary in real-space and one 
    boundary in redshift-space ({\it hybrid}-space).
  \item $C_r$: Selected with both boundaries in real-space.
\end{itemize}  
The real-space and redshift-space boundaries of
Fig.~\ref{fig:sch_phi_real_red} are represented by solid and dotted
lines respectively.  Any auto-correlation of galaxies with this
selection function will essentially be a weighted sum (based on the
amplitude of the selection function) of the auto-correlations of
galaxies within the individual subsamples and the cross-correlations
of galaxies in different subsamples.

In order to investigate the projected clustering of these different
subsamples, we have drawn samples of particles from the HV simulation
(see Section~\ref{sec:hv_sim}), created in top-hat bins of width
100$\hompc$. Sample $A_s$ has top-hat selection boundaries in
redshift-space, sample $B_h$ has one real-space and one redshift-space
boundary, while sample $C_r$ has both boundaries in real-space. These
samples cover the same region of the simulation.

\begin{figure}
  \centering
  \resizebox{0.9\columnwidth}{!}{\includegraphics{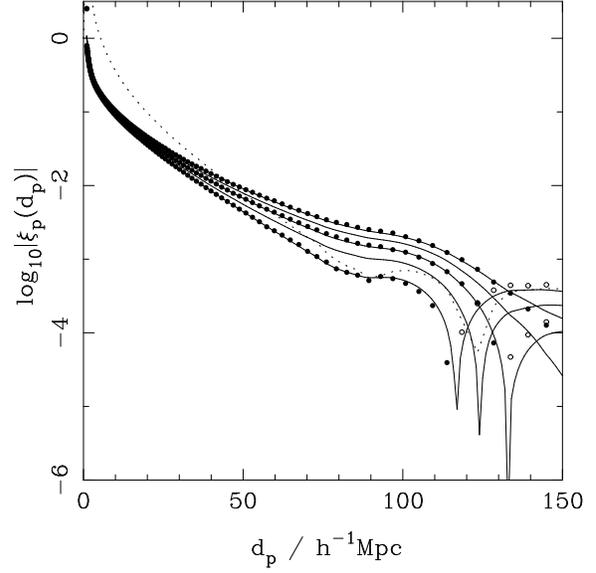}}
  \caption{The average recovered auto-correlation function (solid
    circles) for galaxies from 90 samples drawn from the Hubble Volume
    simulation using three different radial selections, each with
    top-hat width 100 h$^{-1}$Mpc. These are compared against model
    correlation functions calculated for different galaxy samples
    Eq.~(\ref{eq:xi_h}).  The three radial selections are: 1) two
    real-space boundaries (lowest points), which best matches the
    model calculated using the real-space correlation function, 2) two
    redshift space boundaries (highest points), which best matches the
    model calculated using the redshift-space correlation-function
    and, 3) a real-space boundary on one side and a redshift space
    boundary on the other side (points in the middle), which best
    matches the model calculated using the geometric mean of the real-
    and redshift-space correlation functions.}
 \label{fig:autoABC}
\end{figure} 

Fig.~\ref{fig:autoABC} shows the projected auto-correlation functions
for these subsamples.  The measured $\xi_p$ for the $C_r$ and $A_s$
samples are essentially the same as those shown in the top-right panel
of Fig.~\ref{fig:xi_hv_slice}, and just as before they return the
expected real and redshift-space correlation functions calculated via
Eqns.~(\ref{eq:xi_proj_red}) \&~(\ref{eq:xi_mu}).  However, the
hybrid-space correlation function, $\xi_p^h$ of sub-sample $B_h$ has
an amplitude that lies in-between those of the pure real and
redshift-space correlation functions.  We find that we can effectively
model $\xi_p^h$ by assuming the underlying 3D overdensity field has a
correlation function $\xi^h$ given by
\begin{equation}
  \xi^h + 1 = \sqrt{(1+\xi^r)(1+\xi^s)}.
\end{equation}
Note that we are using $\xi^r$ to represent the real-space
3-dimensional correlation function.  As can be seen in
Fig.~\ref{fig:autoABC}, this model is well-matched to the measured
$\xi_p$. The justification for this model is that the multiplicative
boost to the projected density fluctuations ($R$ if we consider that
$\xi=D/R-1$) can be decomposed into multiplicative contributions from
each boundary. Following this model, we should find that the relative
effect of redshift-space distortions on each population, and their
cross-correlations, are simply proportional to the number of
redshift-space boundaries present. If we choose galaxies from a sample
with $n\in\{0,1,2\}$ redshift-space boundaries, and another from a
sample (possibly the same one) with $m\in\{0,1,2\}$ redshift-space
boundaries, then expected correlation function is given by
\begin{equation}  
  \xi^h + 1 = (1+\xi^r)^{1-l/4}(1+\xi^s)^{l/4},
  \label{eq:xi_h}
\end{equation}
where $l=m+n$.

\begin{figure}
\centering
\resizebox{0.9\columnwidth}{!}{\includegraphics{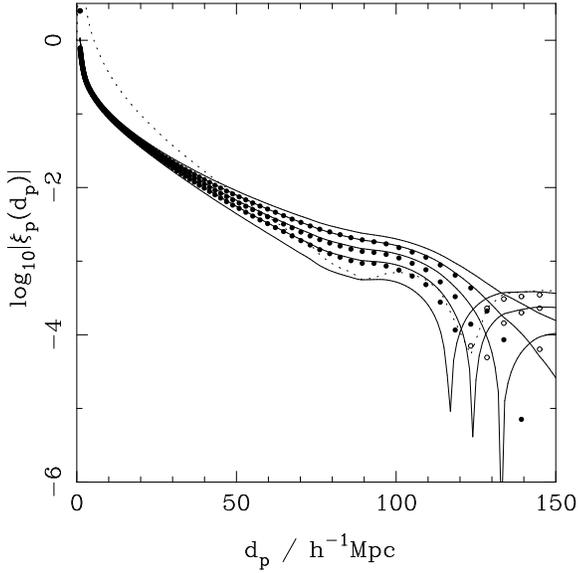}}
\caption{The average measured cross-correlation functions from 90
  radial slices of width 100 h$^{-1}$Mpc in real-space, redshift-space
  or a hybrid with one real-space and one redshift-space boundary,
  each containing $10^6$ galaxies (solid circles). These are compared
  against the model $\xi_p^h$ of Eq.~(\ref{eq:xi_h}) (solid lines),
  for different total numbers of redshift boundaries. The amplitude of
  both model and data correlation functions increase with increasing
  dependence on the redshift-space correlation
  function. \label{fig:crossABC}}
\end{figure} 
Fig.~\ref{fig:crossABC} displays the cross-correlations between the
our three HV subsamples.  As expected, the model calculated using the
appropriate $\xi^h$ from Eq.~(\ref{eq:xi_h}) is the closest match to
the measured cross-correlation in every case. All of the models do
over-predict all three measurements at large scales, but we believe
this is reflective of the error associated with our measurements (one
would expect it to be covariant between each sample as they all sample
the same density field). It is possible that we are seeing effects
caused by the coherence of the boundaries with each other that would be
removed for wider bins, such as those we consider in Section~\ref{sec:des}.

Given a hybrid selection function such as that shown in
Fig.~\ref{fig:sch_phi_real_red}, we must split the sample into
populations where we can assume simple boundary conditions for
each. In fact, we can consider solving the projection equation
(e.g. Eqns.~\ref{eq:xi_proj_real} \&~\ref{eq:xi_proj_red} in real-space
and redshift-space) by Monte-Carlo integration over pairs of radial
galaxy locations. For each pair of locations we can determine the
relative contributions from galaxies in each of the subsamples, and
therefore construct a full model for the correlation function.

\section{Implications for future photometric redshift surveys} 
\label{sec:des}

\begin{figure}
\centering
\resizebox{0.9\columnwidth}{!}{\includegraphics{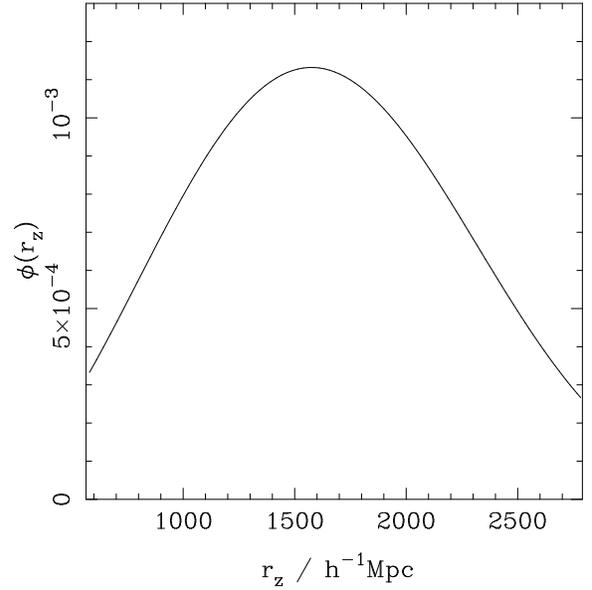}}
\caption{Approximate redshift distribution similar to that expected
  from the Dark Energy Survey. In order to use this distribution of
  galaxies to easily measure cosmological acceleration using projected
  clustering measurements, this population will have to be subdivided
  or binned in redshift. \label{fig:des-phi}}
\end{figure}
\begin{figure}
\centering
\resizebox{0.9\columnwidth}{!}{\includegraphics{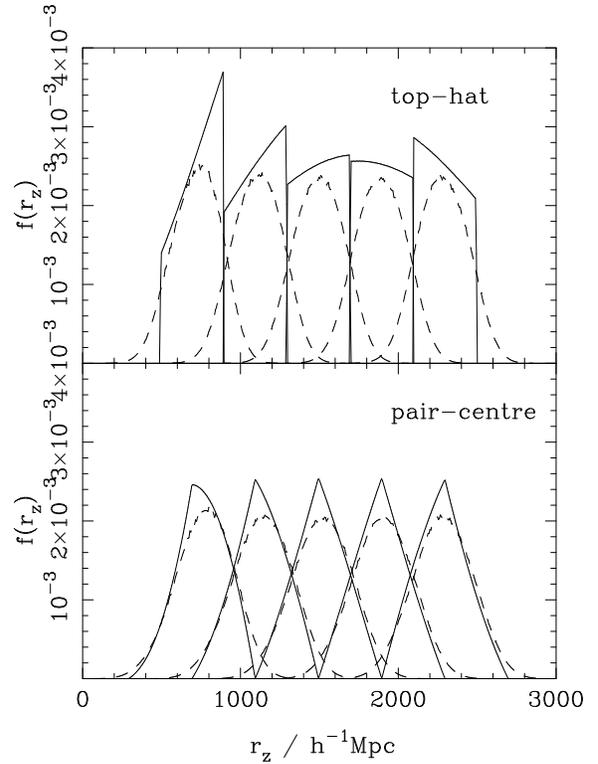}}
\caption{Top panel: normalised radial selection functions for top-hat
  slices of width $400\mpcoh$ created from a DES-like
  distribution. Bottom panel: We also consider bins in radial galaxy
  pair-centre of the same width $400\mpcoh$. While we bin in distances
  derived from photometric redshifts (solid lines), the true
  distribution of radial galaxy distances is shown by the dashed
  lines. \label{fig:photoz-des-slices}}
\end{figure}
A number of extremely wide angle imaging surveys are planned over the
next few years: the Dark Energy Survey (DES), the Panoramic Survey
Telescope \& Rapid Response System (Pan-Starrs) and the Large Synoptic
Survey Telescope (LSST). One goal of these surveys is to constrain the
current acceleration of the Universe. In general, one can hope to use
such surveys to make four measurements of dark energy using
complimentary techniques: cluster counting, BAO, weak lensing and
supernovae. In this paper we consider BAO measurements. For these
experiments, radial distances to galaxies will be estimated from
photometric redshifts, so there will be little information in the
radial direction on the scale of BAO. Consequently, analyses will tend
to rely on making projected galaxy clustering measurements in redshift
slices that are sufficiently narrow to be able to reveal cosmological
acceleration.

In order to assess the effect of redshift-space distortions on such
measurements, we now consider one of these surveys, DES, in more
detail. The DES will use a 500 Mega-pixel camera on the Blanco 4-metre
telescope in Chile to conduct a galaxy survey over a sky area of
5000\,deg$^2$. Multi-band observations using {\it g, r, i} and {\it z}
filters will allow photometric redshifts to be obtained over a range 
$0.2<z<1.4$. The expected redshift distribution of the galaxies will
be approximately \footnote{We thank the DES LSS working group for
  providing this approximation}
\begin{equation} \label{eq:phi}
 \phi_{DES}(z)\propto \left ( \frac{z}{0.5} \right )^2 
    \exp {\left ( -\frac{z}{0.5} \right )^{1.5}},
\end{equation}
after applying approximate survey depths to basic luminosity
functions. This function is plotted in Fig.~\ref{fig:des-phi}.  This
distribution of galaxies will then be sub-divided into bins in order
to assess the evolution of the BAO scale across the survey. As
discussed above, measurements of the projected correlation function
will be affected by redshift-space distortions, which will increase
the signal strength and decrease the importance of BAO features. We
now consider how the choice of binning methodology affects the impact
of redshift-space distortions.

We consider splitting this galaxy distribution into five redshift
slices each of width $400\mpcoh$ for distances estimated from
photometric redshifts, assumed to be Gaussian with
$\sigma_z=0.03(1+z)$. These bins cover radial distances of
$500\to2500\mpcoh$, related to redshifts $z=0.15$ to $z=1.06$
(assuming a flat $\Lambda$CDM cosmology with $\Omega_m=0.25$. The
upper panel of Fig.~\ref{fig:photoz-des-slices} shows the
distributions of galaxies in these slices. The lower panel of
Fig.~\ref{fig:photoz-des-slices} shows the redshift distributions when
we bin the galaxies based on the centre of the radial separation,
calculated from the photometric redshifts. Because we are using
photometric redshifts, there is no way to bin without leaving overlap
in the true radial distributions. Consequently, the top-hat binning
scheme does not provide an obvious advantage over other schemes in
terms of analysing disjoint regions.

\begin{figure*}
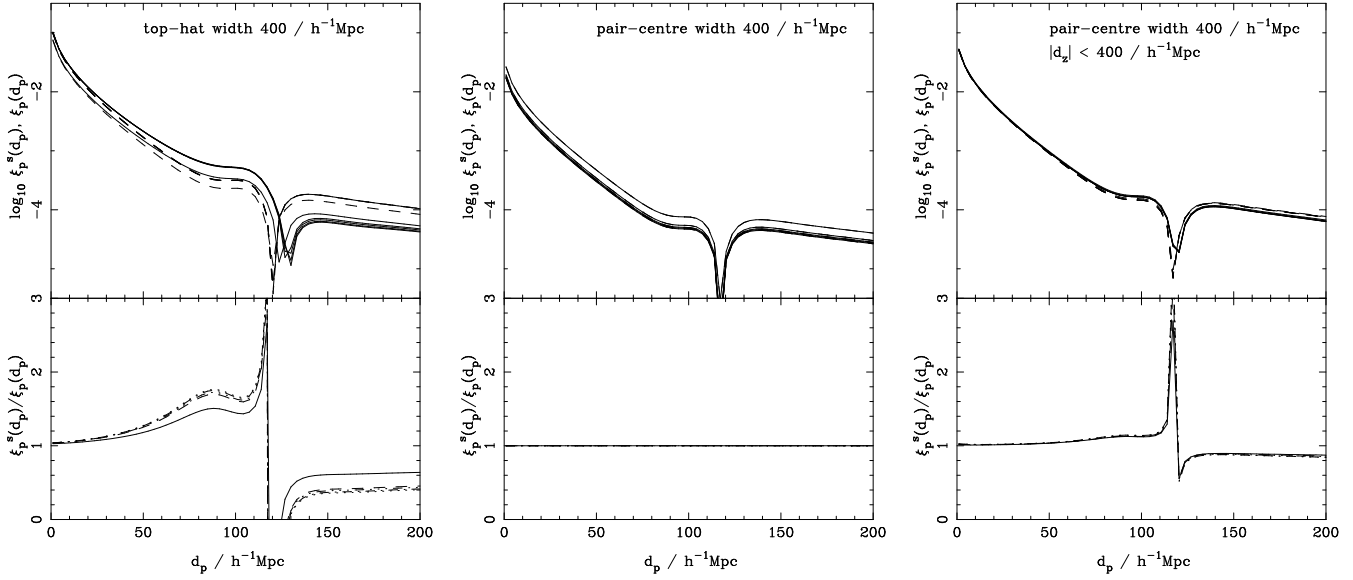

\centering
\centering
\resizebox{0.32\textwidth}{!}{\includegraphics{des_hybrid_win1.ps}}
\hfill
\resizebox{0.32\textwidth}{!}{\includegraphics{des_hybrid_win2.ps}}
\hfill
\resizebox{0.32\textwidth}{!}{\includegraphics{des_hybrid_win3.ps}}
\caption{Top panels: Real-space (dashed lines) and redshift-space
  (solid lines) correlation functions predicted for the 5 radial bins
  drawn from the DES-like selection function, assuming it can be
  treated as a real-space boundary. Bottom panels: The ratio between
  the redshift-space and real-space projected correlation
  function. Here different line styles correspond to different bins:
  in the order of increasing redshift, they are solid, dashed,
  dot-dash, dotted, dot-dot-dash. From left to right: Top-hat bins of
  width $400\mpcoh$ in the radial direction, pair-centre bins of width
  $400\mpcoh$, and constrained pair-centre bins of width $400\mpcoh$,
  including an additional constraint on the radial separation of
  $|d_z|<400\mpcoh$. \label{fig:xi-des}}
\end{figure*}

\begin{figure*}
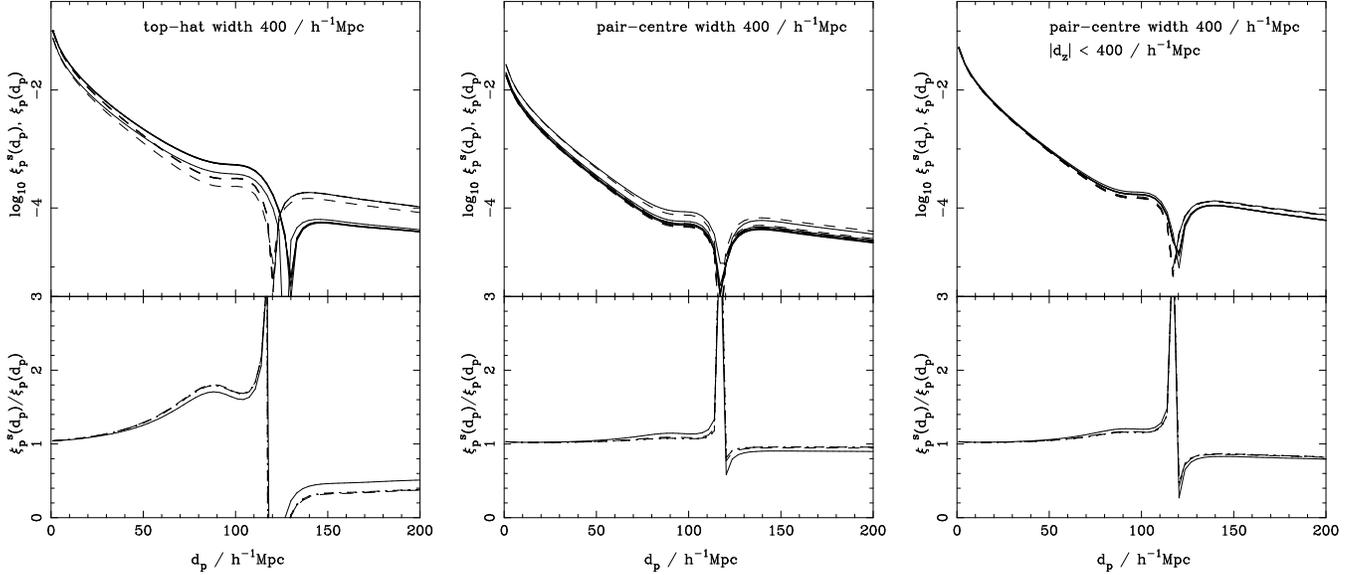

\centering
\centering
\resizebox{0.32\textwidth}{!}{\includegraphics{des_red_win1.ps}}
\hfill
\resizebox{0.32\textwidth}{!}{\includegraphics{des_red_win2.ps}}
\hfill
\resizebox{0.32\textwidth}{!}{\includegraphics{des_red_win3.ps}}
\caption{As Fig.~\ref{fig:xi-des}, only in this case we treat the DES
  selection function as a redshift-space
  boundary. \label{fig:xi-dess}}
\end{figure*}

In light of the discussion in Section~\ref{sec:mag}, we consider both the
case in which $\phi_{DES}$ is treated as a real-space boundary
(results presented in Fig.~\ref{fig:xi-des}), and the case in which it
is treated as a redshift-space boundary (as may be the case when the
slope of $k_{corr}(z)$ is especially large; results presented in
Fig.~\ref{fig:xi-dess}). For the hybrid boundary, we employ the
techniques described in Section~\ref{sec:hybrid} to determine the full form
of the projection. When we treat $\phi_{DES}$ as a redshift-space
boundary, we can simply use Eq.~(\ref{eq:xi_proj_real})
\&~(\ref{eq:xi_proj_red}) to determine $\xi_p$ in real and
redshift-space.

The left-hand panels of Figs.~\ref{fig:xi-des} and~\ref{fig:xi-dess}
show the expected projected correlation functions when a top-hat
binning scheme is applied with width 400$\mpcoh$.  Even for this large
bin width, in every radial bin there is a significant difference
between the result obtained using the redshift-space correlation
function and the real-space correlation function.  The difference is
made clear by observing the ratios between the two, displayed in the
bottom panels.  The ratios are slightly higher in the case where we
treat $\phi_{DES}$ as a redshift-space boundary
(Fig.~\ref{fig:xi-dess}), and the difference between the two treatments
is largest for the lowest redshift bin (which has its selection most
affected by the overall DES selection).  In every case, the ratio is
significant ($\sim$ 1.5) around 100$\mpcoh$ and the shape of the
predicted $\xi_p$ and $\xi_p^s$ measurements differ substantially.

The effects of redshift distortions are completely removed when a
pair-centre binning scheme is employed and the $\phi_{DES}$ boundary
is assumed to be real-space, as made clear in the middle panel of
Fig.~\ref{fig:xi-des}.  Based on the discussion in
Section~\ref{sec:dep_win}, we can simply use
Eq.~(\ref{eq:xi_proj_real}) for both and thus their ratio is
identically 1.  Even when the $\phi_{DES}$ boundary is assumed to be
in redshift-space, as displayed in the middle panel of
Fig.~\ref{fig:xi-dess}, the difference between the redshift-space and
real-space model is considerably smaller than for the top-hat binning.
The right-hand panels of Figs.~\ref{fig:xi-des} and~\ref{fig:xi-dess}
show that even if one applies the constraint that the separation
between pairs be less than $400\mpcoh$ to be included in a constrained
pair-centre bin, redshift space distortions introduce a much smaller
effect than for a top-hat binning scheme.

\section{Discussion}  \label{sec:discussion}

Redshift distortions produce a strong effect on projected clustering
measurements --- one that is far stronger than the redshift-space
distortion effect on the 3D clustering signal for galaxy samples with
low bias and a narrow radial window.  It is clear that redshift
distortion effects must be included when modelling the projected
galaxy clustering in redshift slices.

If we consider the apparent motion of galaxies as we move from real-
to redshift-space, then redshift-space distortions cause an apparent
coherent motion of galaxies into and out of samples. This is true
whether samples have sharp boundaries, or if the selection function
changes more gradually with distance. In fact, we have argued that
such motion does not in itself alter the projected correlation
function --- we would recover the real-space projected correlation
function if we could correct for the movement of the boundary
(i.e. allow for the depth of the survey to change with the
distortions). However, this is not easy to do, although it is
theoretically possible and is an interesting alternative approach. The
effect of redshift-space distortions is due to the redshift-space
boundaries themselves having an angular clustering signal, and their
correlation with the overdensity field. We can alternatively view the
effect from a Lagrangian standpoint, where we have to consider that
the projection does not remove redshift-space effects from the
anisotropic correlation function.

We have used Hubble Volume simulations to show that the projected
correlation function can be modelled most easily by integrating the
redshift-space correlation function over the radial selection
function. Galaxy selection will often be a mix of real and
redshift-space constraints, and we have shown that this can be
modelled by splitting the population into samples that can be
considered to have top-hat windows in either real-space,
redshift-space or a hybrid of the two. In the hybrid situation, the
projected correlation function can be modelled using both the
real-space and redshift-space correlation function over the radial
selection function, and that more complicated selection functions can
be effectively modelled in a similar manner. Prior to this publication,
no-one has considered how these hybrid selection functions affect the
recovered projected clustering signal.

\subsection{Pair-Centre Binning}

We have presented a new measurement technique, {\it pair-centre}
binning, and shown that it minimises the effects of redshift space
distortions.  In this new scheme, we only include galaxies where their
apparent {\it pair-centres} lie within a given radial bin, whereas
traditional methods select pairs where both galaxies lie within the
bin. The new scheme includes individual galaxies that lie {\it
  outside} the traditionally applied top-hat boundaries.  This simple
modification acts to reduce the effect of the coherent movement of
galaxies between slice boundaries on projected correlation function
clustering analyses. It is important to note that this new technique
does not {\it prevent} the movement of galaxies between slices;
redshift-space distortions due to peculiar velocities will always
exist in the radial direction. It simply makes sure that they do not
produce a coherent effect on the measurements.

There are two potential disadvantages of the pair-centre binning
scheme.  One is the fact that the same galaxy may be included in
multiple radial bins --- thus introducing a correlation between radial
bins.  Another is the fact that such a scheme results in necessarily
wider radial bins, which causes the clustering signal to be diluted.
We do not feel that either is a large problem.  Applying the more
traditional top-hat binning scheme to photometric surveys necessarily
results in overlapping radial bins (due to photometric redshift
errors) and there will always be considerable covariance between
radial bins selected with photometric redshifts --- we do not think
that pair-centre binning will make this problem considerably
worse. The dilution effect can be mitigated by imposing a maximum
separation between the pairs included in a pair-centre bin: we
  call this constrained pair-centre binning. As can be seen by
comparing the middle and right-hand panels of Figs.~\ref{fig:xi-des}
and~\ref{fig:xi-dess}, imposing such a constraint increases the
expected signal while not causing a significant change in the effects
of redshift-space distortions.  More detailed studies of these effects
are warranted, but we are confident that the reduction in the redshift
distortion effect we observe when utilising pair-centre binning will
make this scheme considerably preferable to a top-hat binning scheme.

Pair-centre binning completely removes the effect of redshift
distortions when given a uniform galaxy distribution. Such perfect
distributions do not exist --- most galaxy samples selections are
based on an apparent magnitude limit --- and thus realistic radial
distributions of galaxies are more complicated. However, we have
argued that if galaxy samples selected based on an apparent magnitude
limit are cut back so that no galaxies $k$-corrected galaxies are
missing from the sample, then this does not matter: the boundaries of
the bins are either in real-space, or based on pair-centres, neither
of which introduces redshift distortion effects.

We have argued, and it is clear from previous work, that any
interpretation of projected clustering measurements must account for
redshift space distortions. In fact, comparing correlation functions
calculated using different binning schemes might actually prove to
provide a mechanism for measuring the amplitude of the redshift-space
distortions. This is beyond the scope of our current draft, and we
leave this for subsequent work.

\subsection{Future Surveys}

To quantify the effect of redshift-space distortions for future
surveys, we have used the expected radial selection function and
photometric redshift distribution for the Dark Energy Survey to
predict the effect of redshift-space distortions on projected
clustering measurements. This analysis is also relevant to other
planned surveys such as PanStarrs and the LSST, which will have
similar radial selection functions.  We have contrasted two different
types of binning: top-hat --- in which we only allow galaxies between
a given radial bound to enter our sample--- and pair-centre --- in
which we only count galaxy pairs with an average radial position that
lies within our bounds.  For typical bin widths that will be applied
to these surveys, we find that top-hat binning in the radial direction
leaves a strong signal from redshift-space distortions. Using a
pair-centre binning scheme reduces the redshift-space distortion
signal, by as much as 80$\%$ in realistic situations (see
Fig.~\ref{fig:xi-des}) and should therefore allow the measurements to
be more sensitive to the cosmological parameters one wishes to
constrain.

In this analysis, we have only considered the simplified situation
where the redshift-space distortions act along one axis of a Cartesian
basis. However, the arguments we have put forward in favour of
pair-centre binning do not rely on this assumption, and will remain
valid even when wide-angle effects are included in any analysis.

\section{Acknowledgements}

The authors thank the UK Science and Technology Facilities Research
Council for financial support. WJP is also grateful for support from
the Leverhulme Trust and the European Research Council. Simulated data
was calculated and analysed using the COSMOS Altix 3700 supercomputer,
a UK-CCC facility supported by HEFCE and STFC in cooperation with
CGI/Intel.

We thank the DES Large-Scale Structure working group members, and
especially Enrique Gaztanaga, for many helpful discussions. We also
thank the referee for carefully reading our manuscript and providing
excellent suggestions for improvements.

\label{lastpage}


\begin{thebibliography}{99}

  \bibitem[\protect\citeauthoryear{Baldauf et al.}{2009}]{baldauf09}
    Baldauf T., Smith R.E., Seljak U., Mandelbaum R., 2009, 
    [[arXiv:0911.4973]]

  \bibitem[\protect\citeauthoryear{Blake et al.}{2007}]{blake07}
    Blake C., Collister A., Bridle S., Lahav O., 
    2007, MNRAS, 374, 1527

  \bibitem[\protect\citeauthoryear{Cole et al.}{2005}]{cole05}
    Cole S., et al., 2005, MNRAS, 362, 505

  \bibitem[\protect\citeauthoryear{Colless et al.}{2003}]{colless03}
    Colless M., et al., 2003, [[astro-ph/0306581]]

  \bibitem[\protect\citeauthoryear{Eisenstein et al.}{2005}]{eisenstein05}
    Eisenstein D.J., et al., 2005, ApJ, 633, 560
        
  \bibitem[\protect\citeauthoryear{Evrard et al.}{2002}]{evrard02}
    Evrard A.E., et al., 2002 ApJ, 573, 7
    
  \bibitem[\protect\citeauthoryear{Fisher et al.}{1993}]{fisher93}
    Fisher K.B., Scharf C.A., Lahav O.,
    1993, MNRAS, 266, 219
    
  \bibitem[\protect\citeauthoryear{Gaztanaga et al.}{2008}]{gaztanaga08}
    Gaztanaga E., Cabre A., Hui L., 2008, [[arXiv:0807.3551]]

  \bibitem[\protect\citeauthoryear{Hamilton}{1992}]{hamilton92} 
    Hamilton A.J.S., 1992, ApJ, 385, L5

  \bibitem[\protect\citeauthoryear{Hamilton}{1998}]{hamilton98} 
    Hamilton A.J.S., ``Linear redshift distortions: A review'', in ``The
    Evolving Universe'', ed.~D.~Hamilton, pp.~185-275 (Kluwer Academic,
    1998) [[astro-ph/9708102]]
    
  \bibitem[\protect\citeauthoryear{Kaiser}{1987}]{kaiser87}
    Kaiser N., 1987, MNRAS, 227, 1
    
  \bibitem[\protect\citeauthoryear{Padmanabhan et al.}{2007}]{padmanabhan07}
    Padmanabhan N., et al., 2007, MNRAS, 378, 852

  \bibitem[\protect\citeauthoryear{Peebles}{1973}]{peebles73}
    Peebles P.J.E., 1973, ApJ, 185, 413

  \bibitem[\protect\citeauthoryear{Peebles}{1980}]{peebles80}
    Peebles P.J.E., 1980, {\it The Large Scale Structure of the Universe},
    Princeton University Press

  \bibitem[\protect\citeauthoryear{Percival et al.}{2001}]{percival01}
    Percival W.J., et al., 2001, MNRAS, 327, 1297

  \bibitem[\protect\citeauthoryear{Percival et al.}{2007a}]{percival07a}
    Percival W.J., et al., 2007a, ApJ, 657, 51

  \bibitem[\protect\citeauthoryear{Percival et al.}{2007b}]{percival07b} 
    Percival W.J., Cole S., Eisenstein D., Nichol R., Peacock J.A.,
    Pope A., Szalay A., 2007b, MNRAS, 381, 1053

  \bibitem[\protect\citeauthoryear{Percival \& Schafer}{2008}]{percival08} 
    Percival W.J., Schafer B., 2008, MNRAS, 385, L78

  \bibitem[\protect\citeauthoryear{Percival et al.}{2009}]{percival09}
    Percival W.J., et al., 2009, MNRAS submitted, [[arXiv:0907:1660]]

  \bibitem[\protect\citeauthoryear{Perlmutter et al.}{1999}]{perlmutter99} 
    Perlmutter S., et al., 1999, ApJ, 517, 565

  \bibitem[\protect\citeauthoryear{Regos \& Szalay}{1995}]{regos95}
    Regos E., Szalay A.~S.,  1995, MNRAS, 272, 447

  \bibitem[\protect\citeauthoryear{Riess et al.}{1998}]{riess98} 
    Riess A.G., et al., 1998, AJ, 116, 1009

  \bibitem[Ross \& Brunner(2009)]{R09} Ross, A.~J. \& Brunner, 
    R.~J., 2009, MNRAS, 399, 878

  \bibitem[\protect\citeauthoryear{Sanchez et al.}{2009}]{sanchez09}
    Sanchez A.G., Crocce M., Cabre A., Baugh C.M., Gaztanaga E., 2009,
    MNRAS submitted, [[arXiv:0901.2570]]

  \bibitem[\protect\citeauthoryear{Simpson et al.}{2009}]{simpson09}
    Simpson F., Peacock J.A., Simon P., 2009, [[arXiv:0901.3085]]

  \bibitem[\protect\citeauthoryear{York et al.}{2000}]{york00}
    York D.G., et al., 2000, AJ, 120, 1579

\end{thebibliography}
\end{document}